\documentclass[aps,prd,amssymb,eqsecnum,showpacs]{revtex4}
\usepackage{graphicx}
\usepackage{psfrag}

\newcommand{\gz}{\mbox{\em \r{g}\hspace{0.1mm}}}  % Zero order approximation.
\newcommand{\Rz}{\mbox{\em \r{R}}}

\newcommand{\nablaz}{\nabla\hspace{-0.27cm}{}^{\mbox{\r{~}}}{}\hspace{-0.22cm}}
\newcommand{\sigmaz}{\sigma\hspace{-0.23cm}{}^{\mbox{\r{~}}}{}\hspace{-0.12cm}}
\newcommand{\Gammaz}{\Gamma\hspace{-0.20cm}{}^{\mbox{\r{~}}}{}\hspace{-0.18cm}}

\newcommand{\az}{\mbox{\em \r{a}\hspace{0.1mm}}} 

\begin{document}

\title{Geometrization of metric boundary data for Einstein's equations}

\author{Jeffrey Winicour${}^{1,2}$
       }
\affiliation{
${}^{1}$ Department of Physics and Astronomy \\
         University of Pittsburgh, Pittsburgh, PA 15260, USA \\
${}^{2}$ Max-Planck-Institut f\" ur
         Gravitationsphysik, Albert-Einstein-Institut, \\
	 14476 Golm, Germany
	 }

\begin{abstract}

The principle part of Einstein equations in the harmonic gauge
consists of a constrained system of 10 curved space wave equations
for the components of the space-time metric.  A well-posed initial
boundary value problem based upon a new formulation of
constraint-preserving boundary conditions of the Sommerfeld type has
recently been established for such systems. In this paper these
boundary conditions are recast in a geometric form. This serves as a
first step toward their application to other metric formulations of
Einstein's equations.

\end{abstract}


\maketitle

It is extremely satisfying to contribute this article in acknowledgment of the
important influence that J\" urgen Ehlers has had on my career. When I was a
first year graduate student at Syracuse University in the Fall of 1959, my PhD
adviser Peter Bergmann had collected an astonishing percentage of the young
talent engaged in general relativity. Besides the tenured faculty members Art
Komar and Richard Arnowitt, the postdocs and visiting scientists included Roy
Kerr, Wolfgang Kundt, Ted Newman, Roger Penrose, Asher Peres, Ivor Robinson,
Englebert Schucking, Andrej Trautman and Manfried Trumper, as well as J\" urgen.
At the end of that first year, when Bergmann called me to his office to discuss
thesis research topics, from this group of experts it was J\" urgen whom he
invited to join in offering me advice. In response to Bergmann's opening
question ``So, what would you like to work on?'', I suggested singularity
structure. I had recently enjoyed reading the nice paper by Bergmann and his
former student Ray Sachs on singularities in linearized gravitational theory, as
well as the Einstein-Infeld-Hoffman paper on the motion of singularities. J\"
urgen immediately took over the conversation, ``What is your underlying
motivation?'' I naively offered the possibility that singularity structure might
be used to understand elementary particles. I then received a wide ranging
lecture from J\" urgen who carefully explained the conceptual problems
underlying singularities even in electromagnetic theory and why the situation
was much more complicated in general relativity. At that time, a global picture
of the structure of the Schwarzschild singularity, which could not be
anticipated from linearized theory, had just emerged with the help of Kruskal's
extension of the exterior spacetime. The geometric approach of classifying
singularities in terms of the incompleteness of the spacetime manifold had just
begun. The important distinction between spacelike, timelike or null
singularities was not yet recognized. In the following years, an effective
approach to this difficult subject slowly developed from a great deal of effort
by many people. Fortunately for me, I accepted J\" urgen's impromptu lecture as
good advice to steer clear of the subject in my graduate research.

Also at that time, general relativity was a small field going through a
renaissance, which was centered in the United States about Bergmann at Syracuse
and John Wheeler at Princeton. Bergmann's research, which was focused on the
quantization of gravity, had led to several reformulations of Einstein's theory
in terms of Lagrangian, phase space and Hamilton-Jacobi methods. But in all of
these formalisms the same problems associated with dealing with nonlinearity,
identifying the proper observables and handling the constraints posed the same
stumbling blocks against a real physical understanding of quantum gravity that
persist today. However, there were several other current developments which
would have seminal impact on the future of the field: the Kruskal extension of
Schwarzschild spacetime, the Kerr solution, the characteristic description of
gravitational radiation by Bondi and Sachs, followed by Penrose's conformal
version; Penrose's spinor description of gravity, which later led to twistor
theory; and J\" urgen's geometric reformulation of general relativistic
hydrodynamics and thermodynamics, which set the standard for the ensuing
transition in general relativity from the ambiguities of a coordinate dependent,
calculational approach to a geometric approach. The discovery of quasars in the
following years would bring these developments in general relativity to the
attention of astrophysicists, who would become a captive audience to lectures on
the geometry of curved spacetimes, thanks to J\" urgen's help in organizing the
first ``Texas Meeting''.

Throughout my career I have benefited much from J\" urgen's advice, especially
during my two sabbaticals with his group in Garching and my annual visits to the
Albert Einstein Institute, an institute which was created due to his efforts as
founding director. Through my exposure to his standards of clarity and rigor and
his emphasis on a geometric picture, J\" urgen endures as my mentor. I know that
he would much prefer to be the cause of an article of substance rather than
praise, so in that spirit I present the following discussion of the geometrical
aspects of the initial-boundary value problem (IBVP) in general relativity.
Quite some time ago, J\" urgen and his student Saskia Kind~\cite{kind1} treated
the spherically symmetric IBVP for a general relativistic fluid, and they later
applied the work to stellar oscillations in collaboration with Bernd
Schmidt~\cite{kind2}. At that time, little was known about the treatment of
boundaries in general relativity outside of the spherically symmetric case. Here
I treat the vacuum problem in the absence of symmetries. In recent work,
catalyzed by interactions at the Albert Einstein Institute, the well-posedness
of the IBVP for Einstein's equations has been established using harmonic
coordinates~\cite{wpgs}. This puts the IBVP on the same analytic footing as the
Cauchy problem, whose well-posedness was also established using harmonic
coordinates in the classic work of Choquet-Bruhat~\cite{Choquet}. However, the
geometric formulation of the boundary conditions and boundary data for the IBVP
is more complicated than for the Cauchy problem, in which the initial data can
be expressed in terms of the intrinsic metric and extrinsic curvature of the
initial Cauchy hypersurface. In this article I will discuss these geometrical
considerations which underlie the treatment of a boundary.

In the Cauchy problem, initial data on a spacelike hypersurface ${\cal S}_0$ are
extended to a solution in the domain of dependence ${\cal D}({\cal S}_0)$ (which
consists of those points whose past directed characteristics all intersect
${\cal S}_0$).  In the IBVP, data on timelike boundary ${\cal T}$  transverse to
${\cal S}_0$ is used to further extend the solution to the domain of dependence
${\cal D}({\cal S}_0 \cup {\cal T})$.

The IBVP for Einstein's equations has only relatively recently widespread
attention, when its importance to numerical relativity was pointed
out~\cite{stewart}. The first well-posed IBVP was achieved for a tetrad
formulation of Einstein's theory using a first differential order system which
included the tetrad, the connection and the curvature tensor as evolution
fields~\cite{fn}. Subsequently, a well-posed IBVP was formulated for the
harmonic formulation of Einstein's equations as second order wave equations for
the metric~\cite{wpgs}. This extended the classic {\em analytic} treatment of
Choquet-Bruhat~\cite{Choquet} to the well-posedness of the harmonic IBVP. The
initial data for the Cauchy problem has a simple description in terms of the
intrinsic metric and extrinsic curvature of ${\cal S}_0$. The aim of this
article is to present an interpretation of the boundary data in the IBVP in
terms of the geometry of ${\cal T}$. This is not only important for a clearer
understanding of the nature of the gravitational IBVP but also, from a practical
point of view, for the application of boundary conditions in numerical
relativity. The boundary conditions developed in~\cite{wpgs} have been
successfully implemented in a evolution code based upon the harmonic formulation
of Einstein's equations~\cite{cpsombc}. However, much of the numerical work in general
relativity is carried out using other metric formulations, e.g. the BSSN
formulation~\cite{bssn1,bssn2}, where well-posedness of the IBVP  currently
remains an unresolved issue. The geometric formulation of boundary conditions
for the metric presented here is a step in that direction.

I begin with a short review of the Cauchy problem in Sec.~\ref{sec:cauchy},
followed by a discussion of the new difficulties presented by the IBVP and their
analytic resolution using a harmonic formulation in Sec's.~\ref{sec:diff} and
\ref{sec:hibvp}, where the analytic form of the boundary conditions is expressed
in terms of partial derivatives of the harmonic metric. In
Sec's.~\ref{sec:bgeom} and \ref{sec:bg} these are recast in covariant form in
terms of geometric structures intrinsic to the IBVP.

\section{The Cauchy problem}
\label{sec:cauchy}

The initial data for the Cauchy problem can be formulated in a purely
3-dimensional form in terms of the intrinsic metric $h_{\mu\nu}$ and extrinsic
curvature $k_{\mu\nu}$ of the initial Cauchy hypersurface. Here, for notational
simplicity, I will use Greek indices rather loosely to describe either
3-dimensional or 4-dimensional objects. A major notational
complication of the IBVP arises from the $3+1$ decomposition intrinsic to the
Cauchy hypersurfaces and the separate $3+1$ decomposition intrinsic to the
timelike boundary ${\cal T}$. As consistently as possible, I will use lower case
letters, e.g. $h_{\mu\nu}$ and $k_{\mu\nu}$, for geometric objects associated
with the Cauchy hypersurfaces and upper case letters for their counterpart on
the boundary. The Cauchy data are subject to the Hamiltonian and momentum
constraints
\begin{equation}
0={}^{(3)}R +(k^\mu_\mu)^2
      -k_{\mu\nu}k^{\mu\nu} \quad (= 2 G_{\mu\nu}n^\mu n^\nu)
      \label{eq:ham}
\end{equation}
and
\begin{equation}
 0 ={}^{(3)}\nabla_\mu (
  k^\mu_\nu - \delta^\mu_\nu k^\rho_\rho)\quad(= 2 h_\nu^\mu 
    G_{\mu\rho} n^\rho),
    \label{eq:mom}
\end{equation}    
where ${}^{(3)} \nabla_\mu$ is the covariant derivative and ${}^{(3)}R$ is the
curvature scalar associated with $h_{\mu\nu}$. Subject to these constraints,
the Cauchy data determine a solution of Einstein's equations which is unique up
to a diffeomorphism (cf.~\cite{hawkel} for an exposition with many
techniques common to the approach adopted here for the IBVP.)

However, as already hinted by the parenthetical appearance of the Einstein
tensor $G_{\mu\nu}$ in (\ref{eq:ham}) and (\ref{eq:mom}), this disembodied
3-dimensional form of the Cauchy data hides the complexity of the underlying
4-dimensional space-time problem. In order to evolve the data it is necessary
to introduce a foliation of the spacetime by Cauchy hypersurfaces ${\cal S}_t$,
with unit timelike normal $n_\mu$. The evolution of the spacetime metric
$g_{\mu\nu}$ is carried out
along the flow of a timelike vector field $t^\mu$ related
to the normal by the lapse $\alpha$ and shift $\beta^\mu$ according to
$$
    t^\mu= \alpha n^\mu + \beta^\mu \, , \quad  \beta^\mu n_\mu =0.
$$    
In numerical applications, the evolution is coordinatized by a time function
$t$ satisfying ${\cal L}_t t =1$ and $n_\mu= -\alpha \nabla_\mu t$,
and spatial coordinates satisfying ${\cal L}_t x^i =0$, where ${\cal L}_t$ is
the Lie derivative with respect to $t^\mu$.
In this 4-dimensional setting, $h_{\mu\nu}n^\nu=0$  and the spacetime metric
is given by 
$$
    g_{\mu\nu} = -n_\mu n_\nu + h_{\mu\nu}.
$$

An additional complexity, already apparent from the arbitrariness
of the diffeomorphism freedom
in the solution, is that 
Einstein equations do not directly provide a hyperbolic system of
evolution equations.
For that purpose, it is necessary to restrict the gauge freedom. This can be
accomplished by introducing harmonic coordinates, i.e. four solutions
$x^\mu=(t,x^i)$ of the curved space scalar wave equation
$ \Box_g x^\mu  = 0$, so that in these coordinates the metric satisfies
\begin{equation}
{\cal C}^\mu:= \Gamma^\mu = g^{\rho\sigma}\Gamma^\mu_{\rho\sigma}= 
              -\frac{1}{\sqrt {-g}}\partial_\nu\gamma^{\nu\mu} =0,
\label{eq:harmc} 
\end{equation}
where
$\gamma^{\mu\nu}=\sqrt{-g}g^{\mu\nu}$.

This leads to the standard harmonic reduction of the Einstein tensor
\begin{equation}
   E^{\mu\nu}:= G^{\mu\nu} -\nabla^{(\mu}\Gamma^{\nu)} 
       +\frac{1}{2}g^{\mu\nu}\nabla_\rho \Gamma^\rho ,	
\end{equation}
whereby the harmonic conditions (\ref{eq:harmc}) together with the
vacuum Einstein equations give rise to the
quasilinear system of coupled wave equations
\begin{equation}
 0= 2\sqrt{-g} E^{\mu\nu}= g^{\rho\sigma}
       \partial_\rho \partial_\sigma \gamma^{\mu\nu} + N_{\mu\nu} ,
\label{eq:heeq}
\end{equation}    
where $N_{\mu\nu}$ represents terms which do not enter the principle
part.

In the harmonic formulation, the Hamiltonian constraints are satisfied as a
consequence of the harmonic constraints (\ref{eq:harmc}) provided the initial
data  satisfies  $\Gamma^\mu|_{t=0}=0$ and $\partial_t \Gamma^\mu|_{t=0}=0$. The
harmonic conditions also determine $\partial_t \alpha$ and $\partial_t \beta$,
so that the
remaining freedom in the initial gauge data reduce to  $\alpha|_{t=0}$ and
$\beta^\mu|_{t=0}$ We return to the issue of constraint preservation in
Sec.~\ref{sec:constr}. 

In summary, the Cauchy data necessary for determination of a unique space-time
metric consist of $g_{\mu\nu}|_{t=0}=0$ and  $\partial _t g_{\mu\nu}|_{t=0}=0$,
subject to constraints. This 4-dimensional space-time version of the initial
data is referred to as the thin sandwich formulation, as opposed to the
disembodied 3-dimensional version in which  $h_{\mu\nu}|_{t=0}=0$
$k_{\mu\nu}|_{t=0}=0$ are prescribed. The resulting harmonic evolution is not
only unique but the solution depends continuously on the choice of initial data,
i.e. the harmonic Cauchy problem is well-posed~\cite{Choquet}.

\section{Difficulties of the boundary treatment}
\label{sec:diff}

The IBVP has quite different features than the Cauchy problem, as can be
inferred from the simple properties of the flat-space scalar wave equation
$$
 (\partial_t^2 - \delta^{ij}\partial_i \partial_j)\Phi=0 \, \quad x^i=(x,y,z)
$$
in the region $x \le 0$. For a given propagation direction $k^i$, there will be
two characteristics (light rays) $x^i =\pm k^i t$ crossing the Cauchy
hypersurface at $t=0$, but only one characteristic with $k^x>0$ crossing the
boundary at $x=0$. As a result, although the initial Cauchy data consist of
$\Phi|_{t=0}$ and $\partial_t \Phi|_{t=0}$, only half as much boundary data can
be freely prescribed at $x=0$, e.g the Dirichlet data $\partial_t \Phi|_{x=0}$,
or the Neumann data $\partial_x \Phi|_{x=0}$ or the Sommerfeld data $(\partial_t
+ \partial_x) \Phi|_{x=0} $ (based upon the derivative in the outgoing
characteristic direction). In the gravitational case, this  inability to
prescribe both the metric and its normal derivative on a timelike boundary
implies that you cannot freely prescribe both the intrinsic metric of the
boundary and its extrinsic curvature. In terms of the metric components, the
most you can describe is a single quantity, e.g. the Sommerfeld data $K^\mu
\partial_\mu g_{\rho\sigma}$ where $K^\mu$ is an outgoing null direction. Such a
Sommerfeld boundary condition is most beneficial for numerical work since it
allows discretization error to propagate across the boundary (whereas Dirichlet
and Neumann boundary conditions reflect the error and trap it in the grid).

Inability to prescribe both the metric and its normal derivative, complicates
constraint enforcement on the boundary, i.e. the Hamiltonian and momentum
constraints cannot be enforced directly because they couple the metric and its
normal derivative. Instead, the approach in~\cite{wpgs} is to enforce the
harmonic constraints ${\cal C}^\mu=0$ on the boundary and then show that this
(indirectly) leads to the satisfaction of the Hamiltonian and momentum
constraints.

An additional complication is that the domain of dependence of the boundary
${\cal D}({\cal T})$ by itself is empty. Crudely speaking, half of the  past
directed characteristics from each interior point do not meet the boundary. The
Cauchy problem is intrinsically coupled with the boundary problem. At the
intersection of ${\cal S}_0$ and ${\cal T}$ the Cauchy data and boundary data
must be prescribed in a consistent way, or otherwise an artificial shock wave
will be generated. In practice, this is hard to implement without, for
example, using an exact solution in the neighborhood of the intersection.
Moreover, the boundary in general moves relative to the Cauchy hypersurfaces,
i.e. the normal $n_\mu$ to the Cauchy hypersurfaces is not in general tangent to
${\cal T}$. This complicates the geometric relation between the separate $3+1$
decompositions associated with the boundary and the Cauchy foliation.

Furthermore, the boundary does not pick out a unique outgoing null direction at
a given point (but, instead, essentially a half null cone). This complicates the
geometric formulation and interpretation of Sommerfeld boundary data. This is in
addition to the issue of a  physical interpretation of the boundary data. Unless
the boundary is defined by a compact matter distribution, its very existence is
physically {\em artificial}. This is the situation in numerical relativity where
a finite outer boundary is typically introduced even though the most important
numerical output might be the extraction of the gravitational waves that
propagate to infinity. The treatment of such an artificial boundary can
introduce spurious physical effects on the extracted waveform, similar to the
effects arising from initial Cauchy data which contains spurious gravitational
waves. Here I concentrate on the geometrical aspects of the boundary treatment
but the underlying methods can also be used to improve the physical properties
of the treatment of an isolated gravitating system~\cite{isol}, e.g. by the
construction of boundary conditions which lead to asymptotically vanishing
reflection coefficients from a
sufficiently round boundary for increasingly
large radius.

\section{Strongly well-posed constraint-preserving IBVP 
with Sommerfeld boundary conditions}
\label{sec:hibvp}

I begin the discussion of the geometrization of the boundary data with the
analytic formulation of a strongly well-posed treatment of the IBVP for the
Einstein equations in the harmonic gauge based upon~\cite{wpgs,wpe}.
The  harmonic coordinates
$x^\alpha=(t,x^i)$ induce a foliation  of the boundary by $t=const$ surfaces
${\cal B}_t$. This foliation determines a unique future-timelike vector unit
$T^\mu$ which is orthogonal to ${\cal B}_t$ and tangent to the boundary. Here
the metric has the status of an unknown to be determined by solving
(\ref{eq:heeq}), so that other metric related quantities, such as the norm of
$T^\mu$, have similar status. Along with the unit spacelike normal $N^\mu$
which points outward from the boundary, this leads to an orthonormal tetrad
$(T^\mu,N^\mu,Q^\mu.\bar Q^\mu)$ at each point on the boundary, where $Q^\mu$ is
complex null vector tangent to ${\cal B}_t$ with normalization
\begin{equation}
   Q^\mu \bar Q_\mu=2 \, , \quad  Q^\mu Q_\mu=0.
\label{eq:qnorm}
\end{equation}
The tetrad is uniquely determined up to the spin-rotation freedom $Q^\mu
\rightarrow e^{i\theta} Q^\mu$. Uniquely associated with this tetrad
(independent of the choice of $Q^\mu$) are the outgoing and ingoing null vector
fields $K^\mu=T^\mu+N^\mu$ and $L^\mu=T^\mu-N^\mu$, respectively.

The outgoing null vector $K^\mu$ allows us to pose Sommerfeld boundary
conditions. In~\cite{wpgs} it was shown that a properly constructed hierarchy of
Sommerfeld boundary conditions of the type
\begin{equation}
 K^\mu\partial_\mu \gamma^{\rho\sigma} =\text{Sommerfeld data} , \quad
   \gamma^{\mu\nu} =\sqrt{-g} g^{\mu\nu}
\end{equation}
leads to a strongly well-posed IBVP. In addition to the continuous dependence of
the solution on the initial Cauchy data, strong well-posedness implies boundary
stability, i.e. that the solution for the metric and its derivatives can be
estimated in terms of the boundary data~\cite{krlor}.

Certain components of the Sommerfeld data are unconstrained.
These unconstrained data are picked out by the projection tensor~\cite{cpsombc} 
$$ P^\nu_\mu=\delta^\nu_\mu +\frac{1}{2} L_\mu K^\nu, $$
which projects a 1-form $V_\mu$ into the
$(K_\mu,Q_\mu,\bar Q_\mu)$ subspace.

The freely prescribed Sommerfeld data $q^{\rho\sigma}$ are the 6 components
\begin{equation}
     P^\rho_\alpha P^\sigma_\beta K^\mu\partial_\mu \gamma^{\alpha\beta}
     = q^{\rho\sigma},
\label{eq:qfree}
\end{equation} 
where $ q^{\rho\sigma}L_\sigma=0$.    
The remaining boundary conditions enforce the
harmonic constraints ${\cal C}^\nu|_{\cal
T}=0$ by expressing them in the form 
\begin{equation}
  \sqrt{-g}{\cal C}^\nu =\sqrt{-g} \, \Gamma^\nu=-\partial_\rho \gamma^{\rho\nu}= 
   +\frac{1}{2} L_\rho K^\mu\partial_\mu \gamma^{\rho\nu}
             - P^\mu_\rho \partial_\mu  \gamma^{\rho\nu} 
    -P^\rho_\alpha P^\sigma_\beta K^\mu\partial_\mu \gamma^{\alpha\beta}=0,
\label{eq:qconstr}
\end{equation}
which provide Sommerfeld conditions for the remaining $L_\rho$ components of
$K^\mu\partial_\mu \gamma^{\rho\nu}$.

Strong well-posedness results from the hierarchical structure of the boundary
conditions (\ref{eq:qfree}) and (\ref{eq:qconstr}), i.e. the boundary conditions
must form a sequence whose Sommerfeld data only depends on prior members. An
example, which plays an important role in Sec.~\ref{sec:bgeom}, is to prescribe
the  unconstrained components (\ref{eq:qfree}) in the sequence
\begin{eqnarray} 
\frac{1}{2} K_\rho K_\sigma K^\mu  \partial_\mu \gamma^{\rho\sigma} &=&-\sqrt{-g}q_{KK},
 \label{eq:gammkk}         \\
   ( Q_\rho K_\sigma K^\mu -\frac{1}{2}  K_\rho K_\sigma Q^\mu )
      \partial_\mu \gamma^{\rho\sigma}&=-&\sqrt{-g}q_{KQ},
    \label{eq:gammkq}                \\ 
  ( Q_\rho \bar Q_\sigma K^\mu  
    -\frac{1}{2}  K_\rho K_\sigma L^\mu)  \partial_\mu \gamma^{\rho\sigma}
&=&-\sqrt{-g} q_{Q\bar Q} ,
    \label{eq:gammqbq}      \\ 
  ( \frac{1}{2} Q_\rho Q_\sigma K^\mu 
    -  Q_\rho K_\sigma Q^\mu) \partial_\mu  \gamma^{\rho\sigma} 
 &=&-\sqrt{-g} q_{QQ} 
  \label{eq:gammqq}  
\end{eqnarray}     
and then to add on the constrained components (\ref{eq:qconstr}) in
the sequence
\begin{eqnarray}         
   2\sqrt{-g}K_\mu \Gamma^\mu &=&\left( K_\rho L_\sigma K^\mu+  K_\rho K_\sigma L^\mu
            - K_\rho \bar{Q}_\sigma Q^\mu  
            - K_\rho Q_\sigma \bar{Q}^\mu \right)\partial_\mu \gamma^{\rho\sigma} =0,
   \label{eq:Gamk}        \\
  2 \sqrt{-g}Q_\mu \Gamma^\mu &=& \left(  L_\rho Q_\sigma K^\mu + K_\rho Q_\sigma  L^\mu
            - Q_\rho  \bar Q_\sigma  Q^\mu+Q_\rho Q_\sigma  \bar{Q}^\mu \right)
\partial_\mu \gamma^{\rho \sigma} =0,
       \label{eq:Gamq}  \\
  2\sqrt{-g} L_\mu \Gamma^\mu&=& \left( L_\rho L_\sigma  K^\mu+  K_\rho  L_\sigma L^\mu
            - \bar{Q}_\rho L_\sigma Q^\mu
               -  Q_\rho L_\sigma \bar{Q}^\mu \right)\partial_\mu \gamma^{\rho\sigma}=0.
   \label{eq:Gaml}  
\end{eqnarray}    
The sequential structure is determined by the order
$$(K_\rho K_\sigma),\, (K_\rho Q_\sigma),\, (Q_\rho \bar Q_\sigma),\,
(Q_\rho Q_\sigma),\,  (K_\rho L_\sigma),\, (Q_\rho L_\sigma),\,
(L_\rho L_\sigma)
$$
in which the components
of $\partial_\mu \gamma^{\rho \sigma}$ enter into the Sommerfeld boundary
condition for $K^\mu\partial_\mu \gamma^{\rho \sigma}$.

The boundary conditions (\ref{eq:gammkk}) -  (\ref{eq:gammqq}) can be modified
by the addition of lower differential order terms without affecting strong
well-posedness. Such terms can be used to reduce spurious back reflections
from the boundary of an isolated system. (See~\cite{isol} for more details.) In
addition, the boundary conditions (\ref{eq:Gamk}) - (\ref{eq:Gaml}) can
also be modified in the form $$\Gamma^\mu - H^\mu (x.g)=0 $$ to include gauge
forcing terms $H^\mu$ corresponding to a generalized harmonic formulation. Other
examples of Sommerfeld boundary conditions that fit into the hierarchical
structure and that ensure strong well-posedness are considered in~\cite{isol,cpsombc}

The physical content of these boundary conditions can be clarified by
considering the case of a linearized wave incident on a plane boundary. The
first three pieces of free boundary data $(q_{KK}, \, q_{KQ},\,  q_{Q\bar Q})$
are related to the gauge freedom, i.e. they can be set to zero without loss of
generality. The next piece of data $(q_{QQ})$ controls the gravitational
radiation.  The remaining conditions enforce the constraints ${\cal C} ^\mu=0$
on the boundary. This is consistent with the physical expectation that two
pieces (or one complex piece) of data are necessary to describe the two
polarization degrees of freedom of a gravitational wave. More generally, the
boundary has curvature, which necessitates additional nonzero boundary data, as
discussed out in~\cite{isol}. In general, the curvature of the boundary combines
with the radiation modes in a way which cannot be cleanly separated. This is one
of the complications emphasized in the treatment given in~\cite{fn}. It is an
issue I return to in Sec.~\ref{sec:dynamics}. I next consider how to reverse
engineer the above Sommerfeld boundary conditions to clarify their geometrical
content.

\section{Boundary geometry}
\label{sec:bgeom}

I proceed using 4-dimensional notation to describe the
3-dimensional geometrical objects intrinsic to the boundary.
At the most primitive level, these are the intrinsic 3-metric
\begin{equation}
   H_{\mu\nu}=g_{\mu\nu}-N_\mu N_\nu,
\end{equation}
its corresponding 3-connection $D_\mu$
and the extrinsic curvature
\begin{equation}
   {\cal K}_{\mu\nu}=H^\rho_\mu \nabla_\rho N_\nu.
\end{equation}
However, those objects are insufficient to formulate Sommerfeld
boundary conditions because they do not determine a unique outward
null direction.

For that purpose, we introduce an evolution gauge vector field $t^\mu$ tangent
to the boundary. This is in analogy with what is done in a numerical evolution the Cauchy
problem but for the moment we need only define $t^\mu$ on the boundary and not
in the interior. It is not necessary that $t^\mu$ be timelike but its flow must
map, in a future-directed sense, the edge ${\cal B}_0$ (where the
boundary intersects the initial Cauchy hypersurface) into a smooth foliation
${\cal B}_t$ of the boundary. The foliation is parameterized by a time function
$t$ satisfying ${\cal L}_t t =1$, with $t=0$ on  ${\cal B}_0$. (Here ${\cal
L}_t$ represents the Lie derivative with respect to $t^\mu$.) In turn, along
with the intrinsic metric of the boundary, the 2-surfaces ${\cal B}_t$ determine
a unit timelike normal field tangent to the boundary according to $T_\mu =-A
D_\mu t =-A H_\mu^\nu \nabla_\mu t$. The corresponding lapse $A$ and shift
$B^\mu$ intrinsic to the boundary are defined according to $t^\mu= A
T^\mu +B^\mu$.

The intrinsic 2-metric of the boundary foliation is given by 
$$
  Q_{\mu\nu}= Q_{(\mu}\bar Q_{\nu)}=H_{\mu\nu}-T_\mu T_\nu
$$
where we have again introduced a complex null vector $Q^\mu$
with the normalization (\ref{eq:qnorm}). 
The extrinsic curvature of this foliation associated with the
normal $T^\mu$ is
$$
  \kappa_{\mu\nu}= Q^\rho_\mu D_\rho T_\nu.
$$

This structure is sufficient to geometrize the last piece of  free Sommerfeld
data $q_{QQ}$ in the hierarchy (\ref{eq:gammkk}) - (\ref{eq:gammqq}) by relating
it to the extrinsic curvature tensors of ${\cal B}_t$ according to
$$
  Q^\rho Q^\sigma({\cal K}_{\rho\sigma} +\kappa_{\rho\sigma})
=\frac{1}{2}Q^\rho Q^\sigma K^\mu\partial_\mu g_{\rho\sigma} 
    - Q^\rho K^\sigma Q^\mu\partial_\mu g_{\rho\sigma} = q_{QQ} . 
$$
Equivalently, we can express this data in terms of the optical
shear $\sigma$ of the outgoing null congruence determined by $K^\mu$,
\begin{equation}
  2\sigma=Q^\rho Q^\sigma \nabla_\rho K_\sigma  =q_{QQ}. 
  \label{eq:shear}
\end{equation}
In linearized theory, $\sigma$ is the complex data
for an ingoing gravitational wave incident on a plane or spherical boundary.

The free data for the first 3 Sommerfeld conditions (\ref{eq:gammkk}) -
(\ref{eq:gammqbq}) cannot be similarly expressed in terms of the geometry of the
boundary and its tangent space. For example, the first condition, rewritten in
terms of the metric, is
$$
  \frac{1}{2}K^\rho K^\sigma K^\mu \partial_\mu g_{\rho\sigma}= q_{KK}.
$$
Since $K^\mu=T^\mu +N^\mu$, this contains pieces such as $N^\rho N^\sigma N^\mu
\partial_\mu g_{\rho\sigma}$ which can be set to zero by a diffeomorphism
leaving the boundary and its extrinsic curvature intact, e.g. by introducing
Gaussian normal coordinates adapted to the boundary. A geometric formulation of
these boundary conditions requires additional structure to eliminate the effect
of this gauge
freedom. One convenient approach is to introduce  a preferred background
geometry, as described next.

\section{Introduction of a background metric associated with the
initial data}
\label{sec:bg}

Geometrization of the boundary conditions (\ref{eq:gammkk}) - (\ref{eq:gammqbq}) 
can be accomplished by introducing a background 
metric $\gz_{\mu\nu}$ with its associated Christoffel symbols
$\Gammaz^\rho_{\mu\nu}$ and curvature tensor $\Rz^\lambda{}_{\rho\sigma(\mu}$. 
Setting  $f_{\mu\nu} = g_{\mu\nu} - \gz_{\mu\nu}$,
the gauge freedom
in first derivatives of the metric is then fixed
relative to the background by the tensor field
$$
  C^\rho_{\mu\nu}:= \Gamma^\rho_{\mu\nu}-\Gammaz^\rho_{\mu\nu}
   = \frac{1}{2} g^{\rho\sigma}\left( \nablaz_{\mu} f_{\nu\sigma} 
     + \nablaz_{\nu} f_{\mu\sigma} - \nablaz_\sigma f_{\mu\nu} \right).
$$
(In order to avoid confusion, raising and lowering of indices will only
be done with the physical metric $g_{\mu\nu}$ unless otherwise noted.)

The harmonic constraints are now modified to the covariant form
$$
   {\cal C}^\rho := g^{\mu\nu}\left( \Gamma^\rho_{\mu\nu} 
   - \Gammaz^\rho_{\mu\nu} \right) = 0 ,
$$ 
or
$$
 {\cal C}^\rho := g^{\mu\nu}\left( \Gamma^\rho_{\mu\nu} 
   - \Gammaz^\rho_{\mu\nu} \right) - H^\rho =0
$$
if a gauge forcing term $H^\rho(x,g)$ is included. 

Einstein's equations for $ f_{\mu\nu}$ reduce to the quasilinear wave system
\begin{equation}
g^{\rho\sigma}\nablaz_\rho\nablaz_\sigma f_{\mu\nu} 
= 2 g_{\lambda\tau} g^{\rho\sigma} C^\lambda{}_{\mu \rho}C^\tau{}_{\nu\sigma} 
+4 C^\rho{}_{\sigma(\mu} g_{\nu)\lambda} 
    C^\lambda{}_{\rho \tau}g^{\sigma\tau}
    - 2 g^{\rho\sigma}\Rz^\lambda{}_{\rho\sigma(\mu} g_{\nu)\lambda} 
 +2\nabla_{(\mu}H_{\nu)}
 \label{eq:beinst}
 \end{equation}
for which the formulation of a well-posed IBVP 
goes through exactly as before. (See~\cite{isol} for details.)
The null tetrad $(K^\mu, L^\mu, Q^\mu, \bar{Q}^\mu )$ associated with the {\em physical}
metric $g_{\mu\nu}$ is again defined on the boundary in terms of the foliation
constructed by introducing an
evolution field $t^\mu$ tangent to the boundary. This null tetrad is then used to
prescribe a hierarchical set of Sommerfeld boundary conditions. In terms of
$f_{\mu\nu}$ the analogue of  (\ref{eq:gammkk}) - (\ref{eq:Gaml}) take
the  hierarchical form
\begin{eqnarray} 
\frac{1}{2} K^\rho K^\sigma K^\mu  \nablaz_\mu f_{\rho\sigma}  &=& q_K ,
\label{eq:qk} \\
   ( Q^\rho K^\sigma K^\mu  \nablaz_\mu 
    -\frac{1}{2}  K^\rho K^\sigma Q^\mu  \nablaz_\mu) f_{\rho\sigma}&=& q_Q ,
\label{eq:qq} \\ 
  ( L^\rho K^\sigma K^\mu  \nablaz_\mu 
    -\frac{1}{2}  K^\rho K^\sigma L^\mu  \nablaz_\mu) f_{\rho\sigma}
&=& q_L ,
\label{eq:ql} \\ 
   ( \frac{1}{2} Q^\rho Q^\sigma K^\mu \nablaz_\mu 
    -  Q^\rho K^\sigma Q^\mu \nablaz_\mu ) f_{\rho\sigma} 
 &=& q_{QQ} ,
 \label{eq:qqq}   \\
\left( Q^\rho\bar{Q}^\sigma  K^\mu+  K^\rho K^\sigma L^\mu
            - K^\rho \bar{Q}^\sigma Q^\mu  -  K^\rho Q^\sigma \bar{Q}^\mu \right)
\nablaz_\mu f_{\rho\sigma}  &=&  -2K^\mu H_\mu, 
\label{eq:hk}  \\
 \left(  L^\rho Q^\sigma K^\mu + K^\rho Q^\sigma  L^\mu
            - K^\rho L^\sigma  Q^\mu+Q^\rho Q^\sigma  \bar{Q}^\mu \right)
\nablaz_\mu f_{\rho \sigma} &=& -2 Q^\mu H_\mu ,
\label{eq:hq} \\
 \left( L^\rho L^\sigma  K^\mu+  Q^\rho \bar{Q}^\sigma L^\mu
            - \bar{Q}^\rho L^\sigma Q^\mu-  Q^\rho L^\sigma \bar{Q}^\mu \right)
\nablaz_\mu f_{\rho\sigma}&=&  -2 L^\mu H_\mu .
 \label{eq:hl}
\end{eqnarray}
Note that these conditions can be modified by taking linear combinations consistent
with the hierarchy. For example, by subtracting (\ref{eq:qk})
we can replace (\ref{eq:ql}) by
\begin{equation}
      ( N^\rho K^\sigma K^\mu  \nablaz_\mu 
    -\frac{1}{2}  K^\rho K^\sigma N^\mu  \nablaz_\mu) f_{\rho\sigma} = q_N.
\end{equation}

The Sommerfeld boundary
conditions (\ref{eq:qk}) - (\ref{eq:hl}) for $f_{\rho \sigma}$
are now in covariant form and the geometrization the boundary
conditions reduces to eliminating the ambiguity in the
choice of the background metric. We do this by extending the
evolution vector field
$t^\mu$ to a neighborhood of the boundary.
This fixes an  ``evolution gauge'' which allows us to
Lie transport the initial Cauchy data into the future.

The simplest choice is of background metric is
to set
\begin{equation} {\gz}_{\mu\nu}|_{t=0}=g_{\mu\nu}|_{t=0} \, , \quad
  {\cal L}_t  {\gz}_{\mu\nu} =0.
  \label{eq:g0}
\end{equation}  
However, higher order compatibility with the initial Cauchy data can be
incorporated by  setting  
\begin{equation}
 {\gz}_{\mu\nu}|_{t=0}=g_{\mu\nu}|_{t=0} \, , \quad
   {\cal L}_t  {\gz}_{\mu\nu} ={\cal L}_t  g_{\mu\nu}|_{t=0}.
     \label{eq:gt0}
\end{equation}

A further option of (\ref{eq:gt0}) is to to require that the
the background metric is harmonic. This
is most readily expressed in the coordinates $x^\mu =(t,x^i)$ adopted to
the evolution according to 
\begin{equation}
        {\cal L}_t  t =1 \, , \quad {\cal L}_t x^i =0.
\label{eq:adapc}
\end{equation}
Then the densitized background metric
\begin{equation}
  \sqrt{-{\gz}}{\gz}^{\mu\nu}=a^{\mu\nu}+b^{\mu\nu}t + \frac{1}{2} c^{\mu\nu}t^2
  +\frac{1}{6} d^{\mu\nu}t^3,
  \label{eq:harmo}
\end{equation} 
where $a^{\mu\nu}$ and $b^{ij}$ are determined by the initial Cauchy data,
satisfies the harmonicity condition provided
\begin{eqnarray} 
      c^{ij}&=&0\, , \quad  d^{\mu i}=0 \, ,\nonumber \\
   b^{tt}&=&-\partial_i a^{ti}\, , 
     \quad b^{ti}=-\partial_j a^{ij}  \, ,\nonumber \\
   c^{tt}&=&-\partial_i b^{ti} =\partial_i \partial_j a^{ij}  \, , 
       \nonumber\\
 c^{ti}&=&-\partial_j b^{ij} d^{tt}
    =-\partial_i c^{ti}=\partial_i \partial_j b^{ij}\, .\nonumber
\end{eqnarray}
(Here $\gz^{\mu\nu}$, the inverse of $\gz_{\mu\nu}$,  is used to raise
indices.) Other background metrics associated with Lie transport of the initial Cauchy
data could also be used.

For both the alternatives (\ref{eq:gt0}) and ( \ref{eq:harmo}), 
$f_{\mu\nu}=g_{\mu\nu}-{\gz}_{\mu\nu}$ has homogeneous (vanishing) initial
Cauchy data so that they provide the first approximation to a standard iterative
scheme to prove existence of solutions of the quasilinear problem~\cite{krlor}.
From a practical point of view, they also automatically ensure a $C^1$
compatibility between the initial data and boundary data, and thus in a numerical
evolution would reduce spurious high frequency waves arsing at the
edge ${\cal B}_0$.  Such a homogeneous version of the initial value problem can
also be useful in numerical simulations by eliminating the effect of extreme
nonlinear behavior.

In all three cases, (\ref{eq:g0}), (\ref{eq:gt0}) and (\ref{eq:harmo}), the background
${\gz}_{\mu\nu}$ is uniquely
defined by the choice of the evolution vector $t^\mu$. This applies
to any metric based formulation of the Cauchy problem for
Einstein's equations. Note that all such choices of $t^\mu$ are related by diffeomorphism,
so although $t^\mu$ is a critical ingredient it is
a pure gauge object. It does not determine metric information
without further knowledge of the lapse and shift.

Given a choice of evolution vector $t^\mu$, the diffeomorphism freedom in the
solution to the harmonic IBVP can be completely eliminated by tying it to the
adapted coordinates determined by a specific coordinatization of ${\cal S}_0$
according to (\ref{eq:adapc}). Then the results of~\cite{isol} guarantee the
local existence of a unique solution to the generalized harmonic equations
(\ref{eq:beinst}) subject to the boundary conditions (\ref{eq:qk}) -
(\ref{eq:hl}).

\section{Geometrical interpretation of the Sommerfeld data}

A geometrical interpretation of the Sommerfeld boundary data in
(\ref{eq:qk}) - (\ref{eq:qqq}) can
be made by replacing the derivatives $\nablaz_\mu f_{\rho\sigma}$
in terms of derivatives of  $K_\sigma$. The
identity
$(\nabla_\rho - \nablaz_\rho) K_\sigma = -C^\mu_{\rho\sigma}K_\mu
 =\frac{1}{2}K^\mu (\nablaz_\mu f_{\rho\sigma}   
   - \nablaz_\rho f_{\mu\sigma}   -\nablaz_\sigma f_{\mu\rho})$
can be used to rewrite (\ref{eq:qqq}) in terms of    
the optical shear of the boundary foliation,
\begin{equation}
2(\sigma-\sigmaz)=
   Q^\rho Q^\sigma(\nabla_\rho - \nablaz_\rho) K_\sigma = 
   \frac{1}{2} Q^\rho Q^\sigma K^\mu \nablaz_\mu f_{\rho\sigma} 
    -  Q^\rho K^\sigma Q^\mu \nablaz_\mu f_{\rho\sigma} .
 = q_{QQ} .
 \label{eq:sigma}
 \end{equation}
Analogous to (\ref{eq:shear}), this equates the Sommerfeld data $q_{QQ}$ to the
shear of the outgoing null hypersurfaces determined by the boundary foliation,
relative to the background value of the shear. Alternatively, the spin-weight
dependence of this relative shear can be removed by expressing it as the rank
two symmetric tensor
\begin{equation}
    \tilde q_{\mu\nu}:=2(\sigma_{\mu\nu}-\sigmaz_{\mu\nu})
    = (Q^\rho_\mu Q^\sigma_\nu-\frac{1}{2}Q_{\mu\nu})
           (\nabla_\rho - \nablaz_\rho) K_\sigma .
   \label{eq:tsigma}
 \end{equation}

The identity
$$V_\mu  K^\rho(\nabla_\rho - \nablaz_\rho) K^\mu  =
   (V^\rho K^\sigma K^\mu  -\frac {1}{2} K^\rho K^\sigma V^\mu)
           \nablaz_\mu f_{\rho\sigma},
$$  	
which holds for any vector field $V^\mu$, can be used to interpret the boundary
data for the leading conditions 
(\ref{eq:qk}) - (\ref{eq:ql})
in the hierarchy. The geodesic acceleration
of $K^\mu$ with respect to the physical metric and the background metric are
$ a^\mu = K^\nu \nabla_\nu K^\mu$ and $ \az^\mu = K^\nu \nablaz_\nu K^\mu$.
Thus (\ref{eq:qk}) - (\ref{eq:ql}) can be re-expressed as
\begin{eqnarray} 
  K_\mu(a^\mu-\az^\mu)=\frac{1}{2} K^\rho K^\sigma K^\mu 
      \nablaz_\mu f_{\rho\sigma}  &=& q_K ,
      \label{eq:ak}   \\
  Q_\mu(a^\mu-\az^\mu)=(Q^\rho K^\sigma K^\mu
       -\frac {1}{2} K^\rho K^\sigma Q^\mu) \nablaz_\mu f_{\rho\sigma} &=& q_Q    ,
   \label{eq:aq}\\ 
  L_\mu(a^\mu-\az^\mu)=(L^\rho K^\sigma K^\mu  
   -\frac {1}{2} K^\rho K^\sigma L^\mu)
    \nablaz_\mu f_{\rho\sigma}&=& q_L ,
  \label{eq:al}
\end{eqnarray}
where the Sommerfeld data $(q_K,q_Q,q_L)$ is related to the acceleration
by the components of 
\begin{equation}
   q^\mu=(a^\mu-\az^\mu).
   \label{eq:qra}
\end{equation}

Note that neither $a^\mu$ nor  $ \az^\mu$ are uniquely determined by our prior
construction which only defines $K^\mu$ in the tangent space of the boundary.
However, their difference is geometrically and uniquely defined, even though
either $a^\mu$ or  $ \az^\mu$ can be individually set to zero by extending
$K^\mu$ inside the boundary as the tangent vector to the affinely parametrized
null geodesic determined by either the connection $\nabla_\mu$ or
$\nablaz_\mu$, respectively.

Equations (\ref{eq:sigma}) - (\ref{eq:al}) show how the unconstrained 
Sommerfeld data can be described in terms of geometrical objects consisting
of the shear and acceleration of the outgoing null vector $K^\mu$ relative to
their background values. This raises the question whether another geometrical
description can be given which does not introduce a background metric. For the
shear $\sigma$, we already saw that this was possible by using (\ref{eq:shear}).
We can investigate the case of $a^\mu$ by decomposing it into the pieces $a^\mu=
T^\nu \nabla_\nu K^\mu+N^\nu \nabla_\nu K^\mu$. The first piece 
$T^\nu\nabla_\nu K^\mu$ is intrinsic to
the boundary geometry and can be expressed in terms of the extrinsic curvature
of the boundary foliation according to the components
\begin{eqnarray}
    K_\mu T^\nu \nabla_\nu K^\mu &=&0 \,  , \nonumber \\
   Q_\mu T^\nu \nabla_\nu K^\mu &=&Q^\mu T^\nu {\cal K}_{\mu\nu}   
         +Q^\mu \partial_\mu \log A \, , \label{eq:tpiece} \\
    L_\mu T^\nu \nabla_\nu K^\mu &=&  2 T^\mu T^\nu {\cal K}_{\mu\nu} \, , \nonumber
\end{eqnarray}     
where $A$ is the lapse intrinsic to the boundary and ${\cal K}_{\mu\nu} $
is its extrinsic curvature. Consequently, if we were to
prescribe Dirichlet boundary conditions for the metric then the terms $N^\nu
\nabla_\nu K^\mu$ would not enter and the data would be ``boundary intrinsic''. 
No background metric would then be necessary. However, the strong well-posedness
of the IBVP has not been established for the case of Dirichlet boundary
conditions. Furthermore, from a practical viewpoint, homogeneous Dirichlet
boundary conditions are of the reflecting type so that, with the correct
prescription of Dirichlet boundary data, a gravitational wave could propagate
across the boundary but in numerical applications the error would be trapped in
the grid.

It is the $N^\nu \nabla_\nu K^\mu$ piece of $a^\mu$ which requires an extension of
$K^\mu$ inside the boundary. This leads to a more complicated geometric description,
which depends upon the particular way that   $K^\mu$ is extended.
For instance, a boundary defining function $\Phi$ might be introduced
such that $\Phi |_{\cal T}=0$ and
$N_\mu |_{\cal T} =\eta \nabla_\mu \Phi|_{\cal T}$, with
$\eta =(g^{\mu\nu}\nabla_\mu \Phi\nabla_\nu \Phi)^{-1/2}$.
This defines an extension of $N_\mu=\eta \nabla_\mu
\Phi$ to a neighborhood of the boundary, where $\eta$ plays the role of the ``lapse'' for the
boundary defining function. It then follows that
\begin{equation}
  N^\nu \nabla_\nu N_\mu = N_\mu  N^\nu \nabla_\nu \log \eta- \nabla_\mu \log \eta =-D_\mu \log \eta.
\label{eq:npiece}
\end{equation}
Next, by extending $T^\mu$ off the boundary by  the parallel transport $ N^\nu
\nabla_\nu  T^\mu=0$, we have $N^\nu \nabla_\nu K^\mu=N^\nu \nabla_\nu N^\mu$.
Thus, by putting together the pieces (\ref{eq:tpiece}) and (\ref{eq:npiece}),
$a^\mu$ can be expressed in terms of the extrinsic curvature and the lapses $A$
and $\eta$. However, this requires introducing auxiliary  quantities,
i.e. $\Phi$ and the extension of the tetrad vectors $N^\mu$ and $T^\mu$ off the
boundary. Consequently, it does not provide any simpler a geometrical description of the
boundary data than the use of a background metric, except for frame-based
formulations, such as~\cite{fn}, in which the tetrad vectors are evolution variables.

\subsection{Edge data}

The uniqueness of the solution to the harmonic IBVP requires the specification
of the initial values of the lapse and shift, i.e. the relation $t^\mu=\alpha
n^\mu +\beta^\mu$ between the evolution vector $t^\mu$ and the unit timelike
normal  $n^\mu$ to the initial Cauchy hypersurface ${\cal S}_0$. For the pure
Cauchy problem this data is pure gauge information but for the IBVP it contains
geometric information at the edge where ${\cal S}_0$ intersects the boundary
${\cal T}$. This edge data is the hyperbolic angle $\sinh \Theta = n^\mu
N_\mu$ formed by the intersection, which describes the intrinsic motion of the
boundary relative to the Cauchy foliation. The tangency of the evolution vector
$t^\mu$ to the boundary implies that $\sinh \Theta$ is related to the shift by
\begin{equation}
      \sinh \Theta = \frac{1}{\alpha} \beta^\mu N_\mu.
\end{equation}

\subsection{Dynamics of the boundary}
\label{sec:dynamics}

The data for a linearized gravitational wave incident on a plane or spherical 
boundary in a background Minkowski space can be prescribed in terms of the shear
of the outgoing null hypersurfaces. However, if the boundary has a more
dynamical behavior, so that its intrinsic metric and extrinsic curvature
change dynamically in time,
then this description breaks
down. In the generic case, such a boundary gives rise to a dynamically changing
shear even in the absence of linearized curvature. Thus the shear must be
coupled with data determining the dynamics of the boundary in order to
unambiguously  describe the full physical or geometrical  content of the boundary
data.

In the nonlinear theory, the boundary is dynamically traced out by the integral curves of
the unit vector $T^\mu$. (It is also traced out by the integral curves of $t^\mu$
but this description is devoid of metrical content.)
Given the initial value of $T^\mu$, which is supplied
by the initial Cauchy and edge data,  these integral curves can be constructed
in principle from the geodesic curvature
$$ A^\mu = T^\nu \nabla_\nu T^\mu.
$$
(The qualification ``in principle'' is a reminder that the required spacetime
metric and connection are unknowns until the IBVP is solved.) However, only the
normal component of  $N_\mu A^\mu$ enters into determining the dynamics of the
boundary. We immediately have 
$$T_\mu A^\mu=0
$$
and, since $T_\mu=-A D_\mu t$ (where $A$ is
the lapse internal to the boundary and $D_\mu$ its internal connection), we
also have 
$$Q_\mu A^\mu =- Q^\mu T^\nu D_\nu(A D_\mu t) =- Q^\mu T^\nu A D_\mu D_\nu t
=Q^\mu D_\mu \log A.
$$ 
Thus the components of $A^\mu$ tangential to the boundary describe the
freedom corresponding to the lapse in the foliation of the boundary.

The essential piece of data that controls $N_\mu A^\mu$ in the
boundary conditions (\ref{eq:sigma}) - (\ref{eq:al})
is $L_\mu q^\mu=L_\mu(a^\mu-\az^\mu)$. A  straightforward calculation gives
\begin{equation}
  L_\mu q^\mu=-2N_\mu A^\mu + L_\mu N^\nu(\nabla_\nu-\nablaz_\nu)K^\mu
       -L^\mu T^\nu \nablaz_\nu K^\mu,
       \label{eq:NA}
\end{equation}
which is independent of the extension of the tetrad vectors off the boundary.
Since the boundary conditions determine a unique solution of the IBVP,
(\ref{eq:NA}) uniquely determines the boundary values of $N_\mu A^\mu$. Hence,
the dynamical properties of the boundary are controlled in the statement of the
IBVP by the data $L_\mu q^\mu$, although in a very implicit manner.

\section{Constraint preservation}
\label{sec:constr}

The preservation of the harmonic constraints,
$${\cal C}^\rho := g^{\mu\nu}\left( \Gamma^\rho_{\mu\nu} 
   - \Gammaz^\rho_{\mu\nu} \right) -H^\rho(x,g)= 0$$
follows from applying the Bianchi identities to the reduced
evolution system. Consider any formulation of Einstein's
equations  for which the reduced evolution equations $E^{\mu\nu}=0$ take the form
\begin{equation}
     E^{\mu\nu}:= G^{\mu\nu} -\nabla^{(\mu}{\cal C}^{\nu)} 
       +\frac{1}{2}g^{\mu\nu}\nabla_\rho{\cal C}^\rho 
       +A^{\mu\nu}_\sigma {\cal C}^\sigma=0,
       \label{eq:reduced}
\end{equation}
for some smooth coefficients $A^{\mu\nu}_\sigma(x,g,\partial g)$. (This includes
constraint modified versions of the generalized harmonic system.)
The Bianchi identity $\nabla_\mu G^{\mu\nu} =0$ then implies a homogeneous
wave equation for ${\cal C}^\mu$,
\begin{equation}
\nabla^\rho \nabla_\rho \, {\cal C}^\mu +R^\mu_\rho \,{\cal C}^\rho
  -2\nabla_\rho(A^{\mu\rho}_\sigma {\cal C}^\sigma) =0.
  \label{eq:bianchi}
\end{equation}
If the boundary conditions enforce ${\cal C}^\rho |_{\cal T}=0$ and the initial
data enforces  ${\cal C}^\rho |_{{\cal S}_0}=\partial_t {\cal C}^\rho |_{{\cal
S}_0}=0$ then the unique solution of (\ref{eq:bianchi}) is ${\cal C}^\rho=0$. As
a result, the Sommerfeld boundary conditions in the geometrical form
(\ref{eq:tsigma})  - (\ref{eq:al}) along with (\ref{eq:hk}) - (\ref{eq:hl}),
which enforce ${\cal C}^\rho |_{\cal T}=0$,  lead to a well-posed harmonic IBVP
in which the constraints ${\cal C}^\rho=0$ are  satisfied everywhere. In turn,
(\ref{eq:reduced}) implies that the Hamiltonian and momentum constraints
$G^{\mu\nu}n_\nu =0$ are also satisfied.

While these Sommerfeld boundary conditions can be formally applied to any metric
formulation of the reduced Einstein equations, an independent check is necessary
to determine whether the Hamiltonian and momentum constraints are preserved for
formulations which do not explicitly contain an evolution system of the form
(\ref{eq:reduced}). An important example is the BSSN system~\cite{bssn1,bssn2}
which is widely used in numerical simulations of binary black holes.  In current
numerical work, the boundary conditions for BSSN evolution systems are applied
in a naive, homogeneous Sommerfeld form to each evolution variable
(cf.~\cite{bssnsom}). Constraint preservation does not hold for harmonic
evolution with these naive boundary conditions  and cannot be expected to hold
for other systems. The geometric nature of the boundary conditions
(\ref{eq:tsigma})  - (\ref{eq:al}) suggest that they could be applicable to the
BSSN system, although the boundary conditions (\ref{eq:hk}) - (\ref{eq:hl})
enforcing the harmonic constraints would undoubtedly need modification. This
issue deserves further investigation. 

\section{Summary}

Beginning with an analytic description of a strongly well-posed version of the
(generalized) harmonic IBVP, we have shown how the boundary conditions on the
metric may be expressed in a geometric form. The end result can be summarized as
follows. On a manifold ${\cal M}$ with boundaries ${\cal S}_0$ and ${\cal T}$
meeting in an edge ${\cal B}_0$, we introduce an evolution vector field $t^\mu$
whose streamlines are tangent to the boundary and provide a smooth foliation
${\cal S}_t$ of ${\cal M}$ which intersects the boundary in in a smooth
foliation ${\cal B}_t$. Although $t^\mu$ contains no metric information, it
supplies the essential gauge information to (i) introduce a background metric on
${\cal M}$  by the Lie transport of the Cauchy data $g_{\mu\nu}|_{t=0}$ and $
{\cal L}_t  g_{\mu\nu}|_{t=0}$ on ${\cal S}_0$ and (ii) introduce a null tetrad
on ${\cal T}$ which is adapted to the foliation ${\cal B}_t$. Sommerfeld
boundary data is then prescribed for the relative acceleration $q^\mu$, as given
in (\ref{eq:qra}), and the relative shear $\tilde q_{\mu\nu}$, as given in
(\ref{eq:tsigma}), of the outgoing null vector to ${\cal B}_t$ (relative to
their background values). Along with the hyperbolic angle $\Theta$ 
characterizing the initial velocity of  ${\cal T}$ relative to  ${\cal S}_0$,
this data uniquely determines a harmonic spacetime metric (locally in time) by
solving (\ref{eq:beinst}).

These boundary conditions have a hierarchical Sommerfeld form which is
beneficial for numerical application. Although formally they can be applied to
any metric-based evolution system, an  unresolved issue is whether these
boundary conditions, or some modification, are constraint preserving for
hyperbolic reductions of the Einstein equations other than harmonic.

The geometrization of the boundary conditions for the gravitational IBVP has
been presented in 4-dimensional form in which the data is prescribed in terms of
the spacetime metric and its derivatives on ${\cal S}_0$ and ${\cal T}$. A
question of further geometric interest is whether the data can also be presented
in a disembodied 3-dimensional form, as is possible for the Cauchy problem by
prescribing the initial data $h_{\mu\nu}$ and $k_{\mu\nu}$ intrinsic to the
3-manifold ${\cal S}_0$. Such Cauchy data then determines a spacetime metric
solving Einstein's equations which is unique up to a diffeomorphism. If there is
such a disembodied version of the boundary data for the IBVP in terms of the
3-manifolds ${\cal S}_0$ and  ${\cal T}$, and their intersection ${\cal B}_0$,
it does not appear that it can be as simple as in the pure Cauchy problem. In
the treatment given here, the  evolution vector $t^\mu$ plays key roles in both
dealing with the diffeomorphism freedom and describing the boundary data in
geometric form. The introduction of $t^\mu$ allows construction of a
4-dimensional background metric based upon the Cauchy data. A disembodied
version would at the least require introducing $t^\mu$ on ${\cal T}$, along with
the construction of a 3-dimensional background metric on ${\cal T}$. I leave it
as an open question whether an equivalent 3-dimensional version can be given.

%%%%%%%%%%%%%%%%%%%%%%%%%%%%%%%%%%%%%%%%%%%%%%%%%%%%%%%%%%%%%%%%%%%%
\begin{acknowledgments}
%%%%%%%%%%%%%%%%%%%%%%%%%%%%%%%%%%%%%%%%%%%%%%%%%%%%%%%%%%%%%%%%%%%%

This research was supported by NSF grant PH-0553597 to the University of
Pittsburgh. Much of the work has been based upon ideas developed during
previous collaborations with H-O. Kreiss, O. Reula and O. Sarbach. I have also
profited from many discussions with H. Friedrich on the IBVP. Finally, my
thanks to J\" urgen, whose memory continues to be a guiding spirit.

\end{acknowledgments}

\end{document}